\documentclass[reprint,
superscriptaddress,
showpacs,preprintnumbers,
 amsmath,amssymb,
 aps,
 prl,
]{revtex4-1}

\usepackage{color}%not in original tex draft, added to color changes
\definecolor{MyColor}{RGB}{0,0,0}

\usepackage{graphicx}% Include figure files
\usepackage{epstopdf}
\usepackage{placeins}
\usepackage{dcolumn}% Align table columns on decimal point
\usepackage{bm}% bold math
\usepackage{epsfig}
\usepackage{hyperref}% add hypertext capabilities
\begin{document}

\preprint{APS/123-QED}

\title{Scanning tunneling spectroscopy of superconducting LiFeAs single crystals: Evidence for two nodeless energy gaps and coupling to a bosonic mode}% Force line breaks with \\

\author{Shun Chi}
%\affiliation{Department of Physics and Astronomy, University of British Columbia, Vancouver BC, Canada V6T 1Z1}
\author{S. Grothe}
\affiliation{Department of Physics and Astronomy, University of British Columbia, Vancouver BC, Canada V6T 1Z1}
\affiliation{Quantum Matter Institute, University of British Columbia, Vancouver BC, Canada V6T 1Z4}
\author{Ruixing Liang}
\affiliation{Department of Physics and Astronomy, University of British Columbia, Vancouver BC, Canada V6T 1Z1}
\affiliation{Quantum Matter Institute, University of British Columbia, Vancouver BC, Canada V6T 1Z4}
\author{P. Dosanjh}
\affiliation{Department of Physics and Astronomy, University of British Columbia, Vancouver BC, Canada V6T 1Z1}
\affiliation{Quantum Matter Institute, University of British Columbia, Vancouver BC, Canada V6T 1Z4}
\author{W.N. Hardy}
\affiliation{Department of Physics and Astronomy, University of British Columbia, Vancouver BC, Canada V6T 1Z1}
\affiliation{Quantum Matter Institute, University of British Columbia, Vancouver BC, Canada V6T 1Z4}
\author{S. A. Burke}
\affiliation{Department of Physics and Astronomy, University of British Columbia, Vancouver BC, Canada V6T 1Z1}
\affiliation{Quantum Matter Institute, University of British Columbia, Vancouver BC, Canada V6T 1Z4}
\affiliation{Department of Chemistry, University of British Columbia, Vancouver BC, Canada V6T 1Z1}
\author{D. A. Bonn}
\affiliation{Department of Physics and Astronomy, University of British Columbia, Vancouver BC, Canada V6T 1Z1}
\affiliation{Quantum Matter Institute, University of British Columbia, Vancouver BC, Canada V6T 1Z4}
\author{Y. Pennec}
\affiliation{Department of Physics and Astronomy, University of British Columbia, Vancouver BC, Canada V6T 1Z1}
\affiliation{Quantum Matter Institute, University of British Columbia, Vancouver BC, Canada V6T 1Z4}


%\affiliation{
% Department of Physics and Astronomy, University of British Columbia
%\affiliation{
%\affiliation{%
% Department of Physics and Astronomy, University of British Columbia
% %Authors' institution and/or address\\
% %This line break forced with \textbackslash\textbackslash
%}%


\date{\today}% It is always \today, today,

\begin{abstract}
The superconducting compound, LiFeAs, is studied by scanning tunneling microscopy and spectroscopy. A gap map of the unreconstructed surface indicates a high degree of homogeneity in this system. Spectra at 2 K show two nodeless superconducting gaps with $\Delta_1=5.3\pm0.1$ meV and $\Delta_2=2.5\pm0.2$ meV. The gaps close as the temperature is increased to the bulk $T_c$ indicating that the surface accurately represents the bulk. A dip-hump structure is observed below $T_c$ with an energy scale consistent with a magnetic resonance recently reported by inelastic neutron scattering.\\

%As in many other high T$_C$'s a pronounced dip-hump structure is visible in the STS spectra
%and can now be studied a structually and chemically clean system.

%\begin{description}
%\end{description}
\end{abstract}

\maketitle

%\tableofcontents

In the new family of high temperature superconductors, the iron pnictides, a general consensus is emerging in favor of an $s_{\pm}$ symmetry \cite{IIMazin-LaFeAsO,hirschfeld2011gap,Chubukov_review2011} although other paring states such as p-wave  \cite{FCZhang_p_wave,PALee-p-wave} and d-wave have also been suggested \cite{hirschfeld2011gap,Chubukov_review2011}.  A challenge in unambiguously identifying the pairing state in this class of materials is that experimental investigations are occurring against a backdrop of variations in sample purity and quality. In particular many of these compounds have intrinsic limitations due to high defect density from cation doping in the bulk, or structural or electronic reconstructions which complicate surface sensitive investigations \cite{Hoffman-review-2011,Stewart_review}.

A particularly interesting compound among the pnictides is LiFeAs  which is superconducting without cation substitution \cite{LiFeAs_first,LiFeAs_doping_Clarke}.  This potentially places it in the same position that YBa$_2$Cu$_3$O$_{7-x}$ (YBCO) holds in the cuprates \cite{Liang_YBCO}, a stoichiometric superconductor that can be chemically and structurally perfect enough to avoid artifacts arising from disorder.  LiFeAs has the additional advantage of a natural cleaving plane, exposing a non-polar surface that does not undergo reconstruction \cite{LiFeAs_slab_simlation}, making it well suited to surface sensitive spectroscopic studies such as angle resolved photoemission spectroscopy (ARPES) \citep{ARPESBorisenko,ARPESUmezawa} and scanning tunneling microscopy and spectroscopy (STM and STS) \cite{Hess-STM,Davis2012,Hanaguri2012}, much like the cuprate Bi$_2$Sr$_2$CaCu$_2$O$_{8+x}$ (BSCCO) \cite{STM_High_Tc_RevModPhys,Damascelli_ARPES_review}.

In this letter, we show through STM that the surface is clean and unreconstructed.  Through spatially resolved STS measurements, we find that the gap structure is extremely homogeneous, presenting an opportunity to study a clean and well-defined system.  STS acquired at 2K reveal two nodeless gaps, consistent with a multiband $s_{\pm}$ pairing state \cite{IIMazin-LaFeAsO}.  We find that the temperature dependence of the gap is BCS-like, in contrast to the fluctuation driven transition \cite{kamal1994} and pseudogap phase present above $T_c$ in the cuprates \cite{timusk1999}. All of these simplifying features offer a system in which to study a feature LiFeAs does have in common with the cuprates; a pronounced structure above the superconducting gap, indicating strong coupling to boson modes.

\begin{figure}[ht]
%\begin{tabular}{cc}
%\includegraphics[width=0.20\textwidth]{crystal-structure.png}&
\includegraphics[width=0.49\textwidth]{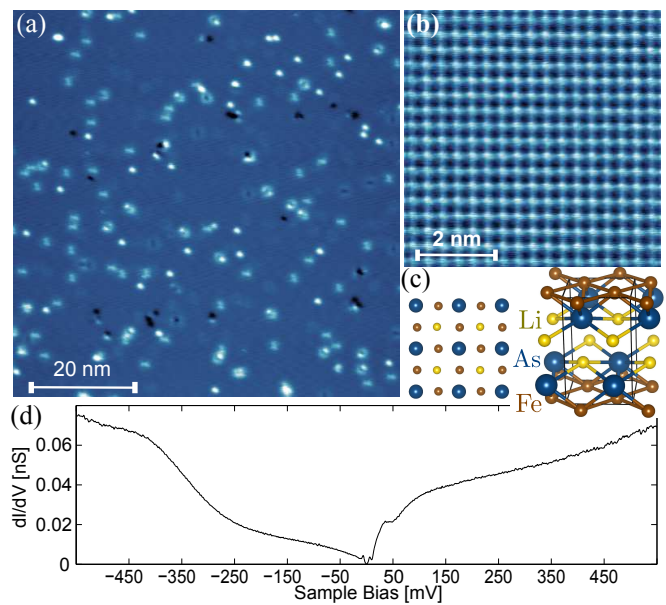}% Here is how to import EPS art
%\end{tabular}
\caption{\label{fig:69-atomic-res} (a) $80 \times 80$ nm$^2$
scanning tunneling topographic image of LiFeAs ($V_B=50$ mV,
$I_T=20$ pA) at T=4.2K and (b) $6.8 \times 6.8$ nm$^2$ atomic resolution
topographic image of LiFeAs ($V_B=40$ mV, $I_T=100$ pA).
(c) Schematic of the crystal structure from 
the cleaved (001) direction
and an edge-on view.
%with grey arrow indicating the anticipated cleavage plane, 
%(see text for details). 
(d) dI/dV spectrum in a range from -550 mV to +550 mV.
}
\end{figure}

%Here we present scanning tunneling spectroscopic measurements of the superconducting gap in this compound. The superconducting properties of the sample observable by STS show a very high degree of spatial homogeneity and a generally BCS-like temperature dependence with a $T_c$ equal to the bulk value.  This allows for an undisturbed analysis of the fundamental properties of superconductivity in LiFeAs.  Spectra reveal at least two nodeless gaps and a pronounced dip-hump feature. The later provides a fresh opportunity to study this widely discussed effect in a clean and well ordered system.

Single crystals of LiFeAs were grown by a self-flux technique.
Li$_{3}$As and FeAs, pre-synthesized from Li (99.9\%), Fe
(99.995\%) and As (99.9999\%), were mixed in a composition of 1:2
and sealed under 0.3 atm Ar. The mixture was heated to 1323 K for
10 hours, then cooled to 1073 K at 4.5 K/hour.
Finally, the samples were additionally annealed at 673 K for 12 hours before being removed from the furnace. 
Single crystals with dimensions up to 2$\times$2$\times$0.2 mm$^3$
were mechanically extracted from the flux. Lattice parameters $a=(3.777\pm0.004)\text{\AA}$\,  and $c =(6.358\pm0.001)
\,\text{\AA}$\ were determined by x-ray diffraction, and $T_{c}^{onset}=17$ K with a transition width %90\%-10\%
of 1K was determined by SQUID magnetometry (see Fig.
\ref{fig:fittings}(c)), consistent with stoichiometric
LiFeAs  \cite{LiFeAs_doping_Clarke}.

All measurements were performed in an ultrahigh vacuum liquid helium temperature STM. 
The electrochemically etched tungsten tips were Ar$^+$ sputtered and thermally annealed at the
beginning of each set of experiments. %and additionally %conditioned by field
%emission over a gold substrate between data sets. 
The samples were
cleaved in-situ at a temperature of $\sim$20 K.
%Friday 16: field emission until 4pm, gap map in the evening
Reproducible datasets were obtained from studies of four different samples, some 
of which have been recleaved and studied with individually prepared tips.
All spectra shown here were acquired by numerical differentiation of the I-V sweep.% allowing for
%a fast data acquisition.% and a low noise due to averaging over several hundrets of individual spectra. 
%Afterwards a Gaussian filter with a width of 10 data points in each direction has been applied. 

%%This might indicate a suitable tip during the preceding spectroscopic measurements.
%This suggests that the presented spectra reflect the intrinsic properties of LiFeAs.

Cleavage most likely occurs between two weakly bonded lithium layers \cite{LiFeAs_slab_simlation}, 
%as indicated by the arrow in Fig. \ref{fig:69-atomic-res}(c),
resulting in a surface that consists of a lithium layer, with arsenic then iron beneath, shown in Fig. \ref{fig:69-atomic-res}(c). A topographic image of an $80\times80$ nm$^2$ area is shown in
Fig. \ref{fig:69-atomic-res}(a), in which several types of defects are visible. The defect density measured from STM images is $0.0020\pm 0.0005$ per LiFeAs formula unit.
% and remained constant in the same area throughout the experiment, indicating these are intrinsic to the crystal or arise from the cleaving process, rather than being adsorbed species. 
Such low defect density further confirms the stoichiometric composition of our LiFeAs crystals. Figure \ref{fig:69-atomic-res}(b) shows an atomic resolution
topographic image with a size of $6.8\times6.8$ nm$^2$ ($V_B=40$ mV, $I_T=100$ pA). The inter-atomic spacing of the square lattice is
($3.74\pm0.03$)\,\AA, which is consistent with the periodicity of
either lithium or arsenic at the surface (see Fig. \ref{fig:69-atomic-res}(c)), but not with the iron lattice.  

Figure \ref{fig:69-atomic-res}(d) shows a dI/dV spectrum in a bias range from $-550$ mV to $550$ mV with a superconducting gap at $E_F$. %This gives insight into the local density of states (LDOS) of LiFeAs. 
%The Fermi energy $E_F$ locates at the dip (trough) of the spectrum, however it is on the cusp of a strongly enhanced empty DOS %that we associate to the bottom of an electron band centred at the M point as observed by ARPES  \cite{ARPESBorisenko, %ARPESUmezawa}. 
%The superconducting gaps, which will be shown in %higher energy resolution later, are very %symmetric even though they are close to the LDOS %enhancement.
The Fermi energy is located in a region of low local density of states (LDOS) enclosed by enhancements below about $-250$ meV and
above $10$ meV. 
The sharp increase of the LDOS just above the Fermi energy reaches a plateau at 35 meV. Similar features have been explained by a surface state, as summarized in  \cite{Hoffman-review-2011}. However, a surface state in LiFeAs is unlikely according to density functional theory \citep{ARPESBorisenko,LiFeAs_slab_simlation}.
The overall U or V shaped background DOS is a generic feature of iron-based superconductors, that has been attributed to their semimetallic character caused by a small overlap of the bottom of electron bands and the top of hole bands close to $E_F$  \cite{Hoffman-review-2011}.  

\begin{figure}[ht]
%\begin{tabular}{cc}
%\includegraphics[width=0.20\textwidth]{crystal-structure.png}&
\includegraphics[width=0.49\textwidth]{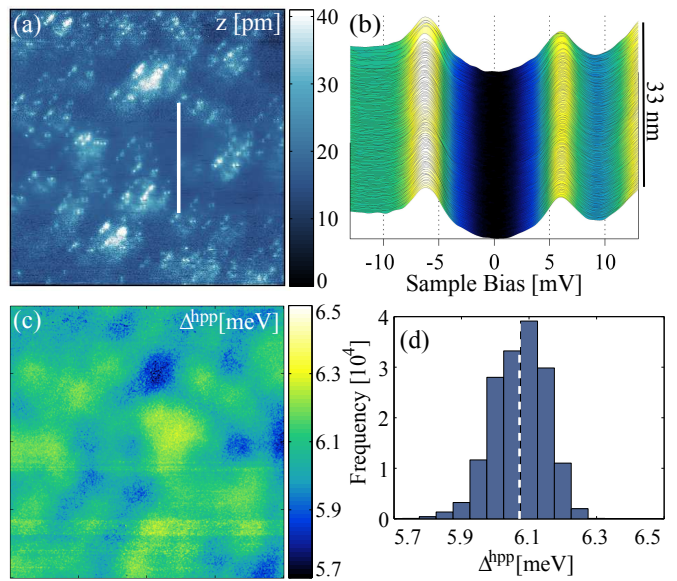}% Here is how to import EPS art
%\end{tabular}
\caption{\label{fig:homog}
(a) $90 \times 90$ nm$^2$ topographic image
($V_B=15$ mV, $I_T=100$ pA) of LiFeAs.
(b) 150 individual $\mathrm dI/\mathrm dV$
spectra (T=4.2K) from the defect-free 33 nm long line marked in
(a). (c) Gap map of the same region in (a). $\Delta^{hpp}$  corresponds to half the energy separation between coherence peaks.
Both, the gap map and the histogram of $\Delta^{hpp}$ shown in (d) reveal
a high degree in homogeneity of the superconducting properties in LiFeAs. 
}
\end{figure}

To gain deeper insight into the spatial homogeneity of the superconducting gap magnitude and in order to study how defects influence the gap magnitude, we recorded $400\times 400$
STS spectra within a bias range of $\pm 15$ mV over an area of $90\times90$ nm$^2$ (see the topography in Fig. \ref{fig:homog}(a)). Figure \ref{fig:homog}(b) shows the 150 measured STS spectra along the 33 nm white line indicated in Fig. \ref{fig:homog}(a), demonstrating a striking homogeneity of the superconducting gap in defect free areas.  A gap map from the whole image is extracted from $\Delta^{hpp}$, which is half the energy separation between apparent coherence peak maxima. $\Delta^{hpp}$ generally overestimates the magnitude of nodeless gaps due to thermal broadening (here T=4.2 K). However, $\Delta^{hpp}$ does allow for an analysis of the spatial variation, independent of any specific model.  
The gap map Fig. \ref{fig:homog}(c) and the corresponding histogram Fig. \ref{fig:homog}(d) indicate a consistent magnitude with a mean value $\overline{\Delta}^{hpp}=6.07$ meV and a small standard deviation $\sigma=0.08$ meV. Much of the variation comes from spectra near defects, where $\Delta^{hpp}$ is reduced to a minimum of 5.7 meV. The homogeneity of the superconducting gap measured in LiFeAs by STS with $\sigma_{\Delta}/\bar{\Delta}\simeq 1.3\%$ is in contrast to most other high temperature superconductors. For comparison, the 122 iron arsenide compound BaFe$_{1.8}$Co$_{0.2}$As$_{2}$ was found to have a less homogeneous gap with $\sigma_{\Delta}/\bar{\Delta}\simeq12\%$  \cite{hoffman-prl-2009}. An extreme example is the much-studied cuprate, BSCCO  \cite{Bi2212_imhomogeneous},where local
inhomogeneities in doping result in strong variation in the gap magnitude. 
%Hence, free from the disorder of local dopants, LiFeAs is one of best compounds to study the mechanism of superconductivity %driving the high-$T_c$ pnicitde superconductors in the clean limit.
Hence, free from surface reconstruction or disorder induced by local dopants, LiFeAs presents an ideal system for surface sensitive spectroscopic investigations.

%\subsection{T dep, fitting}
\begin{figure}[ht]
%\begin{tabular}{cc}
%\includegraphics[width=0.20\textwidth]{crystal-structure.png}&
\includegraphics[width=0.49\textwidth]{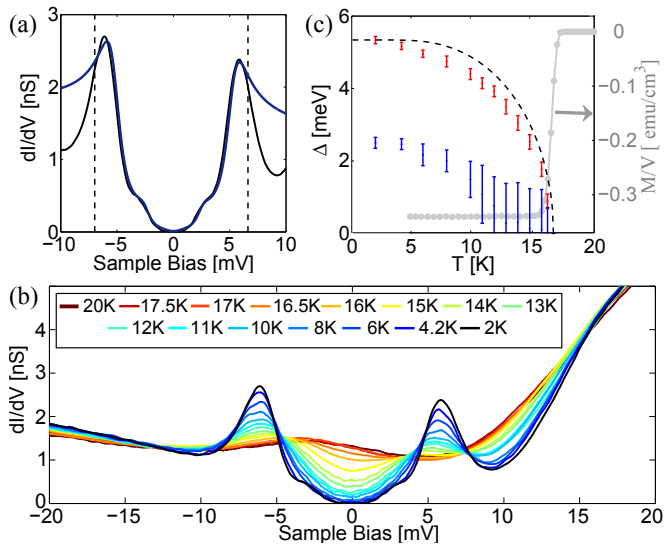}% Here is how to import EPS art
%\end{tabular}
\caption{\label{fig:fittings} 
(a) Two isotropic gaps fit (blue line) on top of measured dI/dV spectra (black line, 2 K). The dashed lines indicate the fitting bias range from -6.8 mV to 6.8 mV. 
(b) Temperature dependence of the superconducting gap measured by STS between 2 and 17 K. 
(c) The large gap determined by isotropic s-wave fits (red error bars) generally follows the temperature dependence predicted by the BCS weak coupling limit (dashed black line).
The development of the smaller gap (blue error bars) is obscured by thermal broadening at elevated temperatures.
The bulk T$_C$ of 17 K probed by SQUID magnetometry with a 1 G magnetic field (gray dots and right y-axis) agrees with the surface critical temperature, demonstrating agreement between surface and bulk properties. 
}
\end{figure} 

Taking a closer look at the STS in the vicinity of the superconducting gap in Fig. \ref{fig:fittings}(a), two nodeless gaps are clearly resolved at 2 K.
The two-gap superconducting excitation spectrum can be fitted within the framework of BCS, where the normalized quasiparticle density of states $\widehat{N}(E)$ of a superconductor is defined as  \cite{Tinkham}
\begin{equation}\label{NomalizeDOS}
\widehat{N}(E)= {\sum_{\vec{k}}}\frac{N_s(E(\vec{k}))}{N_n\left(\sqrt{E(\vec{k})^2-\Delta(\vec{k})^2}\right)}
%\label{eq:normalize}
\end{equation}
{\color{MyColor} 
with the superconducting DOS $N_s$ and normal states DOS $N_n$ being a function of the Cooper
pair energy $E(\vec{k})$ and the single particle energy $\sqrt{E(\vec{k})^2-\Delta(\vec{k})^2}$,
respectively.
If measured by the tunneling method, this can be described by Dynes' formula}  \cite{DynesFormula},
\begin{equation}\label{Tunneling}
    \widehat{N}(eV)=\sum_{\vec{k}}
    \int Re\left[\sum\limits_{i=1}^2
    \frac{w_i\times(E-i\Gamma)\times\frac{\partial f(E-eV)}{\partial
    E}}{\sqrt{(E-i\Gamma)^{2}-|\Delta_i(\vec{k})|^{2}}}
    \right]dE
\end{equation}
where $f(E-eV)$ is
the Fermi-Dirac distribution function and $w_i$ is the weight of the contribution from the $i^{th}$ gap with the constraint of $w_1+w_2=1$ 
\footnote{{\color{MyColor} Because the gap weighting ratio is influenced by the tunneling matrix elements between tip and sample, the relationship between the STS and Eqn. 2 provides only a qualitative assessment of the superconducting DOS.
}}. 
The effective damping term $\Gamma(E)=\alpha E$ is used to properly represent zero DOS at $E_F$ \cite{Davis_Gamma}. 
We fit the 2 K STS over a bias range from -6.8 mV to 6.8 mV under the
assumption that the normal density of states is linear over
the small energy range examined: $N_n(E)=a+b\times E$.

Two different $\Delta(\vec{k})$ superconducting gap models were considered. One consists of two isotropic gaps, yielding $\Delta_1 = 5.33\pm 0.10$ meV, $\Delta_2 = 2.50\pm 0.15$ meV, $w_1$=0.89
and energy dependent $\Gamma=0.13\times E$. The other consists of two anisotropic gaps with four fold symmetry as observed by ARPES \cite{ARPESUmezawa}, yielding $\Delta_1$ = $5.33 \times (1 + 0.09 \times \cos(4\theta))\pm 0.1$ meV ($\Delta_1^{max}=5.8$ meV), $\Delta_2$ = $2.50 \times (1 + 0.20 \times \cos(4\theta'))\pm 0.20$ meV,  $w_1$=0.87 and $\Gamma=0.10\times E$. 
Both gap models fit the 2 K STS very well within the fitting range (see for example the two-isotropic-gap model fit in Fig. \ref{fig:fittings}(b)) {\color{MyColor}and give gaps that are consistent with recently reported values obtained from STS \cite{Davis2012,Hanaguri2012}, giving} reduced gaps of $\frac{2\Delta_1}{k_BT_c}=7.3$ and $\frac{2\Delta_2}{k_BT_c}=3.4$. 
However, both of the fits clearly fail to represent the measured spectra outside the fitting range due to additional above gap features, which will be discussed further below.   

The gap magnitudes obtained from fitting agree well with the largest and the smallest of the four gaps determined by ARPES \cite{ARPESUmezawa}. Tunneling into the two electron pockets, containing the other two gaps, is expected to be strongly suppressed because of the larger in-plane momentum $|\vec{k}_{||}|$  \cite{STM_theory,Hoffman-review-2011}.
The small anisotropy factors obtained from the fit to two anisotropic gaps also agree well with ARPES results \cite{ARPESUmezawa}, reinforcing the consistency of the two surface sensitive measurements in LiFeAs. 
However, unlike the gap shapes previously reported by STS \cite{Hess-STM}, the gaps shown in Fig. \ref{fig:fittings}(a) and (b) are fully open and symmetric even at elevated temperatures.

Fig. \ref{fig:fittings}(b) shows the temperature dependence of the STS spectra from 2 K to 20 K in the same defect-free region. Each spectrum is the average of 16 spectra acquired from a $4\times4$ nm$^2$ area. The superconducting gaps are visible up to 16.5 K and disappear at 17 K, manifesting the same $T_c$ seen in susceptibility.
Based on the simpler isotropic gap assumption, which adequately represents the gap size given the small anisotropy factors found, the temperature dependence of the gap amplitudes was extracted, and is plotted with the Meissner transition measured by SQUID in Fig. \ref{fig:fittings}(c).  The temperature dependence of $\Delta_1$ follows BCS theory and the gap closes at the bulk $T_c$ value.
%changed after 1st review:

These results and the apparent absence of a surface state or electronic reconstruction \citep{ARPESBorisenko,LiFeAs_slab_simlation}
suggest that the surface behavior echoes the bulk.

%We use the simpler two isotropic gap structure to determine the magnitudes of the gaps at higher temperatures with fixed $w$=0.89 (both models produce very similar results). The temperature dependence of the gap magnitudes determined by fitting is shown together with the Meissner transition in Fig. \ref{fig:fittings}(c). The temperature dependence of $\Delta_1$ generally follows BCS theory, and the gap closes at the bulk $T_c$.
\begin{figure}[h]
%\begin{tabular}{cc}
%\includegraphics[width=0.20\textwidth]{crystal-structure.png}&
\includegraphics[width=0.49\textwidth]{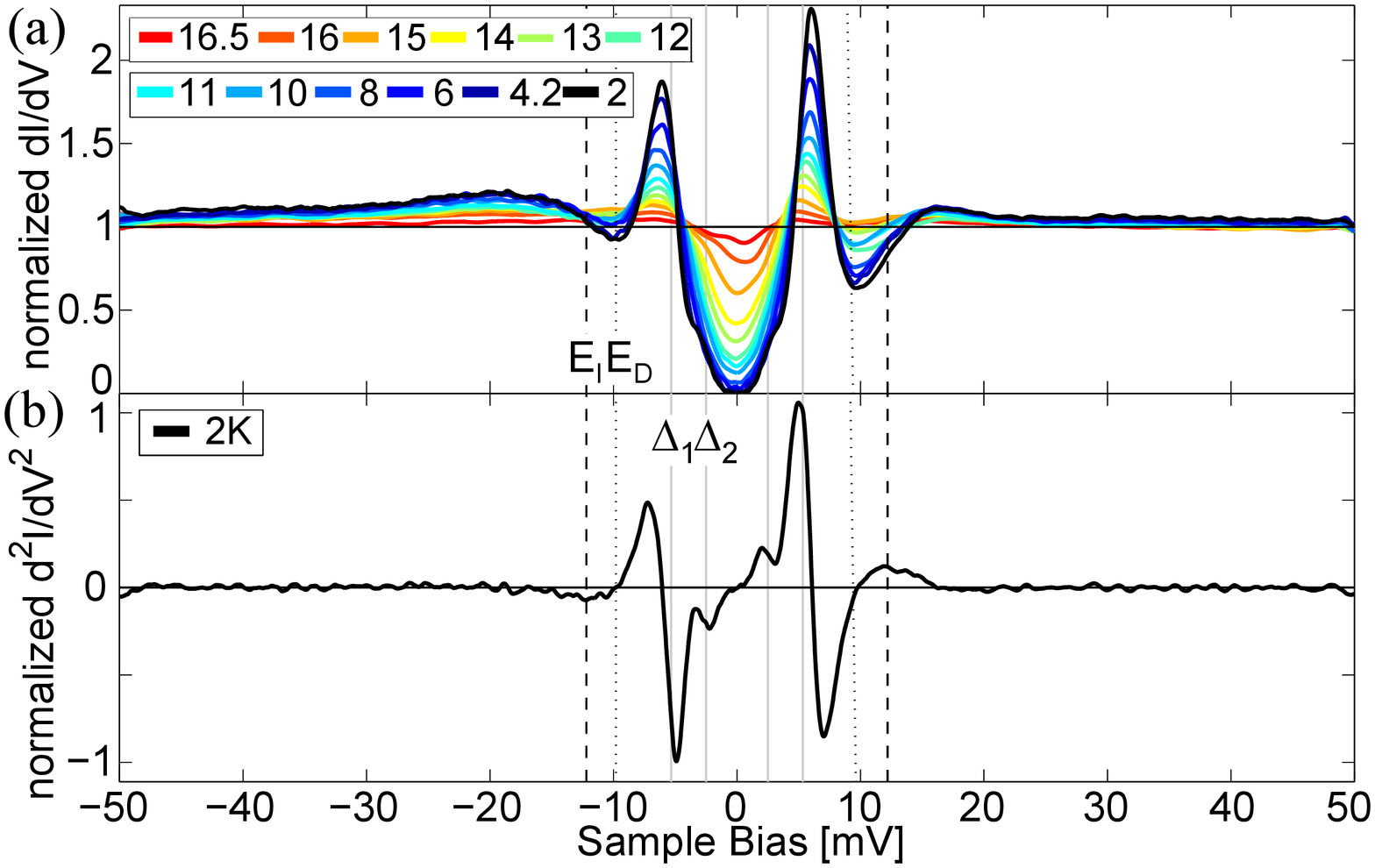}% Here is how to import EPS art
%\end{tabular}
\caption{\label{fig:fig4} 
a) Normalized STS in a temperature range between 2 and 16.5 K.
   A dip-hump feature is clearly visible at 2 K but decays with increasing temperature. 
b) The normalized second derivative of $I(V)$ at 2 K reveals the position of the main features.
   The dip in the normalized $dI/dV$ is at $E_D\approx \pm 10$ mV and the inflection point/kink in $dI/dV$ is at 
   $E_I\approx \pm 12$ mV (dashed line).
   %The characteristic energy distance between $\Delta_1$ (solid line) and the dip-hump kink of about 
   %7 meV is highlighted. 
}
\end{figure}

To more carefully examine the structure surrounding the superconducting gap, the STS below $T_c$ are normalized by the normal state conductance, $dI/dV(V)|_{17K}$ according to 
Eq. \ref{NomalizeDOS}, shown in Fig. \ref{fig:fig4}(a), and the second derivative was calculated numerically, shown in Fig \ref{fig:fig4}(b). Here, features higher in energy than the coherence peaks are clearly visible, diminishing as the temperature approaches $T_c$.  These features consist of a relatively well-defined dip below the normal state conductance, followed by a broad hump.  These  can be characterized by three different energies: $E_D\simeq$10meV, $E_I\simeq$12meV, and $E_H\sim$15-20meV, the energy of the dip, inflection point between dip and hump, and the hump respectively.
%Added after 1st review:

%Both coherence peak and dip-hump feature are more pronounced for positive bias, an asymmetry reminiscent of the cuprates \cite{dewilde1998}. 
%However, such an asymmetry in the present data can be induced by the normalization using the 17 K spectrum marked by the thermally broadended sharp increase of the DOS above 10 meV. %The hump feature at about -4 meV in the normal state might also be responsible for the observed asymmetry.
While the apparent bias asymmetry in the coherence peaks and dip-hump features are reminiscent of the cuprates \cite{dewilde1998}, this may also arise from the normalization to the 17K spectrum which exhibits a hump at around -5meV and a steep rise at positive bias, both of which are thermally broadened compared to the low temperature spectra.

The features observed in LiFeAs, characterized by a dip at $\sim2\Delta$ from $E_F$ followed by a broader hump,
bear striking resemblance to those observed in the cuprates \cite{dewilde1998, zasadzinski2001, pasupathy2008electronic, STM_High_Tc_RevModPhys}.
{\color{MyColor} 
In past studies on superconducting cuprate materials, these features have been attributed to several different origins due to the large variety of competing effects at similar energy scales.
These include inelastic tunneling effects \cite{Lee2006, Pilgram2006},
band structure effects \cite{Kordyuk2002},
though most attribute these spectra features to either pairing \cite{dewilde1998, Eschrig2000, zasadzinski2001, Yeh2001, Levy2008, Ahmadi2011} or non-pairing boson interactions \cite{Johnston2010}.
Other explanations are based on the pseudogap observed in the cuprates \cite{Sacks2006, Gabovich2007}, but these are likely absent in the iron arsenides\cite{Hoffman-review-2011}.
In our data, the reduction of the tunneling conductance below the normal state indicates that inelastic tunneling effects are not responsible for these features, and band structure effects are unlikely in our spectra since the features disappear above $T_c$.}

Thus, we turn our attention to possible boson interactions.
In the well-established framework of phonon-based pairing in an s-wave superconductor  \cite{Scalapino1966}, the energy dependence of the gap leads to an initial peak or shoulder outside the quasiparticle coherence peaks due to increased pairing strength, followed by a dip as the interaction switches from attractive to repulsive at the mode energy. Our spectra do not show this initial peak or shoulder.  Regardless, features caused by coupling to a bosonic mode are expected to appear at the mode energy shifted by the gap ($\Omega+\Delta$)  \cite{dewilde1998,zasadzinski2003}.  Given the differences between our spectra and the classic phonon coupling case, combined with the lack of strong features in the phonon spectrum of LiFeAs below $\sim$14meV  \cite{Li2011phonon,Um2012}, it seems unlikely that the features we observe arise from phonon mediated pairing. 
Spin fluctuation mediated superconductivity has been suggested in the pnictides \cite{IIMazin-LaFeAsO}, a theory also supported by STS data of SmFeAsO$_{1-x}$F$_x$ \cite{Fischer_SmFeAsO2009}. 
Indeed, recent reports from neutron scattering have indicated a broad magnetic excitation peaked around 5-8meV  \cite{LiFeAs_Neutron_spinfluc,LiFeAs_Neutron_spinflucPRL}, corresponding well with the energy of the dip shifted by the large gap, $E_D-\Delta_1\simeq$5meV.  %The appearance of these features at an energy corresponding to magnetic excitations points to a need for the extension of Eliashberg strong coupling theory to a multiband $s_{\pm}$ gap symmetry and possible coupling to magnetic excitations. 
%Added after 1st review:

The cuprates and the arsenides thus share a couple of features that suggest a common origin for the dip-hump.  Both have a mode developing in the spin fluctuation spectrum when the superconducting gap appears below T$_c$ \cite{ timusk1999, Stewart_review}.  Both also have considerable damping of this mode due to non-zero density of states; in the cuprates this is due to nodes in the d-wave gap and in the arsenides {\color{MyColor}this can arise from anisotropy as well as a finite density of states once above the energy of the} small gaps associated with the multiband nature of the s$_{\pm}$ state.  This may reasonably result in similar spectral shapes, however unambiguous assignment of the spectral features will require a proper microscopic theory including the pairing symmetry of LiFeAs. The energy scale of the features found here also draw a parallel to the cuprates. Zasadzinski et al. showed a clear proportionality of the boson mode energy $\Omega$ with $T_c$, that also agreed in magnitude with the magnetic resonance from inelastic neutron scattering for a wide range of $T_c$ from overdoped to underdoped BSCCO\cite{zasadzinski2001}. They found that $\Omega/\Delta\approx 1$ for optimally doped samples and generally that $\Omega\approx 5k_BT_c$, confirmed for a wide range of cuprate superconductors \cite{Hüfner2008}.  The feature observed here for LiFeAs similarly sits at $\Omega/\Delta\approx 1$ and lies close to the $5k_BT_c=7$ meV, indicating a similar underlying mechanism.
%This has been confirmed for a wide range of cuprate superconductors as summarized in \cite{Hüfner2008} and also agrees well with the values obtained here for LiFeAs, where
%$\Omega/\Delta_1\approx 1$ and where $\Omega$ is close to $5k_BT_c=7$ meV.

%\subsection{Conclusion}
The characteristics of LiFeAs presented here demonstrate that this material provides a comparatively simple system in which to study high-$T_c$ superconductivity. In stark contrast to the best cuprate materials, the superconducting gap remains remarkably homogeneous over large areas in LiFeAs.  Additionally, the presentation of a non-polar cleaved surface, that does not reconstruct and accurately represents the bulk properties, makes it ideal for surface sensitive studies.  Although this material shows multiple superconducting gaps, they are without nodes, and exhibit BCS-like temperature dependence.  Yet, this material shows all the signs of strong coupling with a relatively large reduced gap of 7.3, and strong above gap features corresponding closely in energy with a magnetic resonance recently reported.  The sum of these features offers a clearer view into the quasiparticle phenomenology and should serve as a testbed for new understanding of superconductivity.
\begin{acknowledgments} 
The authors would like to thank George Sawatzky, Giorgio Levy, Peter Hirschfeld and Steve Johnston for several helpful conversations. This work was supported by the Canadian Institute for Advanced Research, the Canada Foundation
for Innovation, and the Natural Sciences and Engineering Research Council of Canada.
\end{acknowledgments}

\nocite{*}
\bibliographystyle{apsrev4-1}
%\bibliography{LiFeAs-PRL-March02_shun}
\bibliography{LiFeAs-PRL-June21}% Produces the bibliography via BibTeX.

\FloatBarrier
\end{document}